\documentclass[showpacs,twocolumn,prb]{revtex4}
\usepackage{amssymb}
\usepackage{graphicx}

\begin{document}

\title{Topological Insulators at Room Temperature}
\author{Haijun Zhang$^1$, Chao-Xing Liu$^2$, Xiao-Liang Qi$^3$, Xi Dai$^1$, 
\\ Zhong Fang$^1$,  Shou-Cheng Zhang$^3$ }
\affiliation{$^{1}$Beijing National Laboratory for Condensed Matter
Physics, and Institute of Physics, Chinese Academy of Sciences,
Beijing 100190, China\\
$^{2}$Center for Advanced Study, Tsinghua University, Beijing, 
100084, China
$^{3}$Department of Physics, McCullough Building, Stanford University, 
Stanford, CA 94305-4045}
\date{\today}

\begin{abstract}
Topological insulators are new states of quantum matter with
surface states protected by the time-reversal symmetry. In this
work, we perform first-principle electronic structure calculations
for $Sb_2Te_3$, $Sb_2Se_3$, $Bi_2Te_3$ and $Bi_2Se_3$ crystals. Our
calculations predict that $Sb_2Te_3$, $Bi_2Te_3$ and $Bi_2Se_3$ are
topological insulators, while $Sb_2Se_3$ is not. In particular,
$Bi_2Se_3$ has a topologically non-trivial energy gap of $0.3 eV$,
suitable for room temperature applications. We present a simple and
unified continuum model which captures the salient topological
features of this class of materials. These topological insulators
have robust surface states consisting of a single Dirac cone at the
$\Gamma$ point.
\end{abstract}

\pacs{}
\maketitle

Recently, the subject of time reversal invariant topological
insulators has attracted great attention in condensed matter
physics\cite{day2008,kane2005A,bernevig2006a,bernevig2006d,koenig2007,fu2007a,moore2007,roy2006a,qi2008,hsieh2008}.
Topological states of quantum matter are defined and described by
the corresponding topological terms in quantum field theory. For
example, the quantum Hall effect is described by the topological
Chern-Simons term\cite{zhang1992}. On the other hand, the
electromagnetic response of three dimensional (3D) topological
insulators are described by the topological $\theta$ term of the
form $ S_\theta=\frac{\theta}{2\pi}\frac{\alpha}{2\pi} \int d^3xdt
{\bf E \cdot B}$, where ${\bf E}$ and ${\bf B}$ are the conventional
electromagnetic fields and $\alpha$ is the fine structure constant.
For a periodic system, all physical quantities are invariant under
the shift of the parameter $\theta$ by integer multiples of $2\pi$.
Therefore, all time reversal insulators, including strongly
correlated and disordered ones, fall into two disconnected classes,
described\cite{qi2008} either by $\theta=0$ or by $\theta=\pi$.
Topological insulators are defined by $\theta=\pi$, and this term
physically described the topological magneto-electric
effect\cite{qi2008}. Topological insulators have surface or edge
states with an odd numbers of gapless Dirac points.

The 2D topological insulator displaying the quantum spin Hall (QSH)
effect was first predicted for the HgTe quantum
wells\cite{bernevig2006d}. Recently, the edge state transport has
been experimentally observed in this system\cite{koenig2007}. The
electronic states of the 2D HgTe quantum wells are well described by
a $2+1$ dimensional Dirac equation where the mass term is
continuously tunable by the thickness of the quantum well. Beyond a
critical thickness, the Dirac mass term of the 2D quantum well
changes sign from being positive to negative, and a pair of gapless
helical edge states appear inside the bulk energy gap. This
microscopic mechanism for obtaining topological insulators by
inverting the bulk Dirac gap spectrum can also be generalized to
other 2D and 3D systems. The guiding principle is to search for
insulators where the conduction and the valence bands have the
opposite parity, and a ``band inversion" occurs when the strength of
some parameter, say the spin-orbit coupling, is tuned. For systems
with inversion symmetry, a method based on the parity eigenvalues of
band states at time reversal invariant points can be
applied\cite{fu2007a}. Based on this analysis, the $Bi_xSb_{1-x}$
alloy has been predicted to be a topological insulator for a small
range of $x$, and recently, surface states with an odd number of
crossings at the fermi energy has been observed in angle-resolved
photo-emission spectroscopy (ARPES) experiments\cite{hsieh2008}.

Since $Bi_xSb_{1-x}$ is an alloy with random substitutional
disorder, its electronic structures and dispersion relations are
only defined within the mean field, or the coherent potential
approximation (CPA). Its surface states are also extremely complex,
with as many as five or possibly more dispersion branches, which are
not easily describable by simple theoretical models. Alloys also
tend to have impurity bands inside the nominal bulk energy gap,
which could overlap with the surface states. Given the importance of
topological insulators as new states of quantum matter, it is
important to search for material systems which are stoimetric
crystals with well defined electronic structures, preferably with
simple surface states, and describable by simple theoretical models.
In this work we focus on layered, stoimetric crystals $Sb_2Te_3$,
$Sb_2Se_3$, $Bi_2Te_3$ and $Bi_2Se_3$. Our theoretical calculations
predict that $Sb_2Te_3$, $Bi_2Te_3$ and $Bi_2Se_3$ are topological
insulators while $Sb_2Se_3$ is not. Most importantly, our theory
predicts that $Bi_2Se_3$ has a topologically non-trivial energy gap
of $0.3eV$, therefore, {\bf it is a topological insulator at room
temperature}. The topological surface states for these crystals are
extremely simple, described by a single gapless Dirac cone at the
${\bf k}=0$ $\Gamma$ point. We also propose a simple and unified
continuum model which capture the salient topological features of
this class of materials. In this precise sense, this class of 3D
topological insulators share the great simplicity of the 2D
topological insulators realized in the HgTe quantum wells.

{\it Band structure and parity analysis}. $Bi_2Se_3$, $Bi_2Te_3$,
$Sb_2Te_3$, and $Sb_2Se_3$ share the same rhombohedral crystal
structure with the space group $D^5_{3d}$ ($R\bar{3}m$) with five
atoms in one unit cell. We take $Bi_2Se_3$
as an example and show its crystal structure in Fig. 1a, 
which has layered structures with triangle lattice within one layer.
It has a trigonal axis (three fold rotation symmetry), defined as z
axis, a binary axis (two fold rotation symmetry), defined as x axis,
and a bisectrix axis (in the reflection plane), defined as y axis.
The material consists of five-atom layers arranged along $z$
direction, known as quintuple layers. Each quintuple layer consists
of five atoms with two equivalent $Se$
atoms (denoted as $Se1$ and $Se1'$ in Fig. 1b), 
 two equivalent $Bi$ atoms
(denoted as $Bi1$ and $Bi1'$ in Fig. 1b), and a third $Se$ atom
(denoted as $Se2$ in Fig. 1b). The coupling is strong between two
atomic layers within one quintuple layer but much weaker,
predominantly of the van der Waals type, between two quintuple
layers. The primitive lattice vectors ${\bf t}_{1,2,3}$ and
rhombohedral unit cells are shown in Fig. 1(a).
$Se2$ site plays the role of inversion center and under inversion
operation, $Bi1$ is changed to $Bi1'$ and $Se1$ is changed to
$Se1'$. The existence of inversion symmetry enable us to construct
eigenstates with definite parity for this system.

\begin{figure}
   \begin{center}
      \includegraphics[width=3.5in]{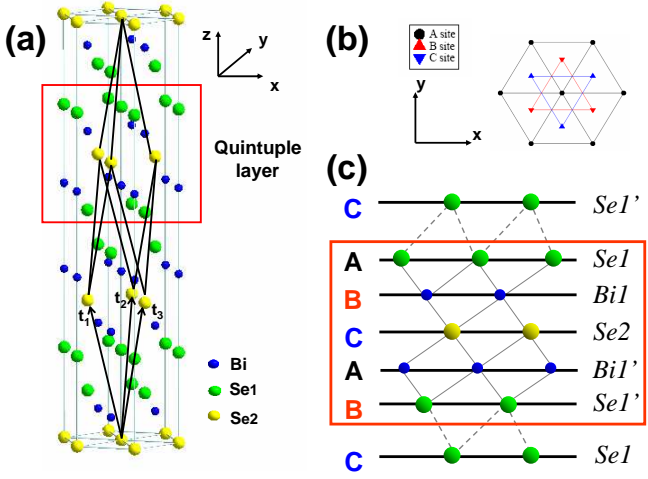}
    \end{center}
    \caption{ {\bf Crystal Structure} (a) Crystal structure of $Bi_2Se_3$ with
     three primitive lattice vectors denoted as $\vec{t}_{1,2,3}$. A quintuple
     layer with $Se1-Bi1-Se2-Bi1'-Se1'$ is indicated by the red box.
     (b) Top view along the $z$ direction. Triangle
     lattice in one quintuple layer has three different positions, denoted as A, B and C.
     (c) Side view of the quintuple layer structure. Along z direction, the stacking order of $Se$ and $Bi$ atomic layers is the
     sequence $\cdots-C(Se1')-A(Se1)-B(Bi1)-C(Se2)-A(Bi1')-B(Se1')-C(Se1)-\cdots$.
     $Se1$ ($Bi1$) layer can be related to $Se1'$ ($Bi1'$) layer by an inversion
     operation of which $Se2$ atoms play the role of inversion center. }
    \label{fig:crystal}
\end{figure}

{\it Ab initio} calculations for $Sb_2Te_3$, $Sb_2Se_3$, $Bi_2Te_3$
and $Bi_2Se_3$ are carried out in the framework of
PBE-type\cite{perdew1996} generalized gradient approximation(GGA) of
the density functional theory (DFT)\cite{hk1964,ks1965}. BSTATE
package\cite{fang2002} with plane-wave pseudo-potential method is
used with $\bf k$-point grid taken as $10\times10\times10$ and the
kinetic energy cut-off fixed to 340eV. For $Sb_2Te_3$, $Bi_2Te_3$
and $Bi_2Se_3$, the lattice constants are chosen from
experiments\cite{sbte1998}, while for $Sb_2Se_3$, the lattice
parameters are optimized in the self-consistent calculation for
rhombohedral crystal structure ($a=4.076$\AA, $c=29.830$\AA), due to
the lack of experiment data.

\begin{figure}
   \begin{center}
\includegraphics[angle=90,width=3.5in]{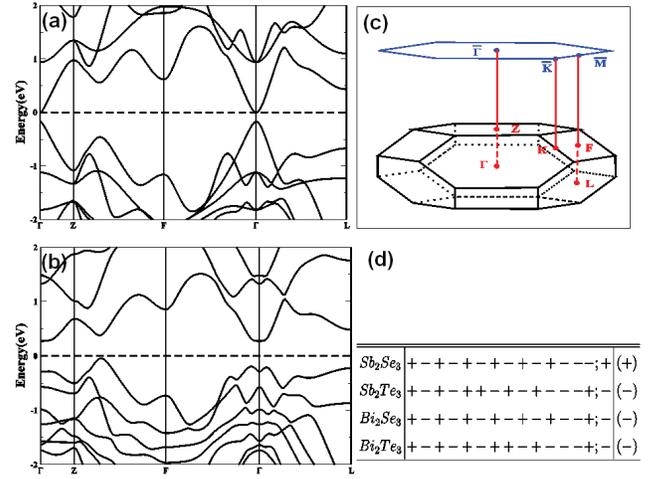}
    \end{center}
    \caption{{\bf band structure, Brillouin zone and parity eigenvalues.}
    Band structure for $Bi_2Se_3$ without spin-orbit
    coupling(SOC) (a) and with SOC (b). The dashed line indicates
    Fermi level.  (c) BZ for $Bi_2Se_3$
    with space group $R\overline{3}m$. The four inequivalent time-reversal invariant
    points are
     $\Gamma(0,0,0)$,L$(\pi,0,0)$, F$(\pi,\pi,0)$ and Z$(\pi,\pi,\pi)$.
     The blue hexagon shows the 2D BZ of projected (1,1,1) surface, in which the high-symmetry {\bf k}
     points $\overline{\Gamma}$, $\overline{\textrm{K}}$ and $\overline{\textrm{M}}$
     are labeled. (d) The parity of the band at $\Gamma$ point for the four materials
    $Sb_2Te_3$, $Sb_2Se_3$, $Bi_2Se_3$ and $Bi_2Te_3$. Here we show the parities of fourteen
    occupied bands, including five s bands and nine p bands, and the lowest unoccupied band.
    The product of the parities for the fourteen occupied bands is given
    in the bracket on the right of each row.
     }
    \label{fig:band}
\end{figure}

Our results are consistent with the previous
calculations\cite{mishra1997,larson2006}. In particular, we note
that $Bi_2Se_3$ has an energy gap about $0.3$eV, which agrees well
with the experimental data (about
$0.2-0.3$eV)\cite{black1957,moose1956}. In the following, we take
the band structure of $Bi_2Se_3$ as an example. Fig. 2 (a) and (b)
show the band structure of $Bi_2Se_3$ without spin-orbit coupling
(SOC) and with SOC, respectively. By comparing the two figures one
can see clearly that the only qualitative change induced by turning
on SOC is an anti-crossing feature around $\Gamma$ point, which thus
indicates an inversion between the conduction band and valence band
due to SOC effect, suggesting $Bi_2Se_3$ to be a topological
insulator. To firmly establish the topological nature of this
material, we follow the method proposed by Fu and Kane\cite{fu2007a}
and calculate the product of the parities of the Bloch wavefunction
for the occupied bands at all the time-reversal invariant momenta
$\Gamma,F,L,Z$ in Brillouin zone (BZ). As expected, we find that at
$\Gamma$ point the parity of one occupied band is changed upon
turning on SOC, while the parity remains unchanged for all occupied
bands at other momenta $F,L,Z$. Since the system without SOC is
guaranteed to be a trivial insulator, we conclude that $Bi_2Se_3$ is
a strong topological insulator. The same calculation is performed
for the other three materials, from which we find that $Sb_2Te_3$
and $Bi_2Te_3$ are also strong topological insulators, and
$Sb_2Se_3$ is a trivial insulator. The parity eigenvalues of the
highest $14$ bands below the fermi level and the first conduction
band at $\Gamma$ point are listed in Fig. 2 (d). From this table we
can see that the product of parities of occupied bands at $\Gamma$
point changes from the trivial material $Sb_2Se_3$ to the three
non-trivial materials, due to an exchange of the highest occupied
state and the lowest unoccupied state. This agrees with our earlier
analysis that an inversion between the conduction band and valence
band occurs at $\Gamma$ point.

To get a better understanding of the inversion and the parity
exchange, we start from the atomic energy levels and consider the
effect of crystal field splitting and spin-orbit coupling to the
energy eigenvalues at $\Gamma$ point, which is summarized
schematically in three stages (I), (II) and (III) in Fig. 3 (a).
Since the states near Fermi surface are mainly coming from $p$
orbitals, we will neglect the effect of $s$ orbitals and starting
from the atomic $p$ orbitals of $Bi$ ($6s^26p^3$) and $Se$
($4s^24p^4$). In stage (I), we consider the chemical bonding between
$Bi$ and $Se$ atoms within a quintuple layer, which is the largest
energy scale in the current problem. First we can recombine the
orbitals in a single unit cell according to their parity, which
results in three states (two odd one even) from each $Se$ $p$
orbital and two states (one odd one even) from each $Bi$ $p$
orbital. The formation of chemical bonding hybridize the states on
$Bi$ and $Se$ atoms, thus push down all the $Se$ states and lift up
all the $Bi$ states. In Fig. 3 (a), these five hybridized states are
labeled as $\left|P1_{x,y,z}^\pm\right\rangle$,
$\left|P2_{x,y,z}^{\pm}\right\rangle$ and
$\left|P0_{x,y,z}^{-}\right\rangle$, where the superscripts $+,-$
stand for the parity of the corresponding states. In stage (II), we
consider the effect of the crystal field splitting between different
$p$ orbitals. According to the point group symmetry, the $p_z$
orbital is split from $p_x$ and $p_y$ orbitals while the latter two
remain degenerate. After this splitting, the energy levels closest
to the Fermi energy turn out to be the $p_z$ levels
$\left|P1^+_z\right\rangle$ and $\left|P2^-_z\right\rangle$. In the
last stage (III), we take into account the effect of SOC. The atomic
SOC Hamiltonian is given by $H_{so}=\lambda \vec{l}\cdot\vec{S}$,
with $l,S$ the orbital and spin angular momentum, and $\lambda$ the
SOC parameter. The SOC Hamiltonian mixes spin and orbital angular
momenta while preserving the total angular momentum, which thus
leads to a level repulsion between
$\left|P1_z^+,\uparrow\right\rangle$ and
$\left|P1_{x+iy}^+,\downarrow\right\rangle$, and similar
combinations. Consequently, the
$\left|P1_z^+,\uparrow(\downarrow)\right\rangle$ state is pushed
down by the SOC effect and the
$\left|P2_z^-,\uparrow(\downarrow)\right\rangle$ state is pushed up.
If the SOC is large enough ($\lambda>\lambda_c$), the order of these
two levels is reversed. To see this inversion process explicitly, we
also calculate the energy levels $\left|P1_z^+\right\rangle$ and
$\left|P2_z^-\right\rangle$ for a model Hamiltonian of $Bi_2Se_3$
with artificially rescaled atomic SOC parameters
$\lambda(Bi)=x\lambda_0(Bi), \lambda(Se)=x\lambda_0(Se)$, as shown
in Fig. 3 (b). Here $\lambda_0(Bi)=1.25{\rm eV}$ and
$\lambda_0(Se)=0.22{\rm eV}$ are the realistic value of $Bi$ and
$Se$ atomic SOC parameters, respectively.\cite{wittel1974} From Fig.
3 (b) one can see clearly that a level crossing occurs between
$\left|P1_z^+\right\rangle$ and $\left|P2_z^-\right\rangle$ when the
SOC is about $60\%$ of the realistic value.
Since these two levels have opposite parity, the inversion between
them drives the system into a topological insulator phase.
Therefore, the mechanism for the 3D topological insulator in this
system is exactly analogous to the mechanism in the 2D topological
insulator of $HgTe$. In summary, through the analysis above we find
that $Bi_2Se_3$ is topologically nontrivial due to the inversion
between two $p_z$ orbitals with opposite parity at $\Gamma$ point.
Similar analysis can be carried out on the other three materials,
from which we see that $Sb_2Te_3$ and $Bi_2Te_3$ are qualitatively
the same as $Bi_2Se_3$, while the SOC of $Sb_2Te_3$ is not strong
enough to induce such an inversion.

\begin{figure}
   \begin{center}
    \includegraphics[width=3.5in]{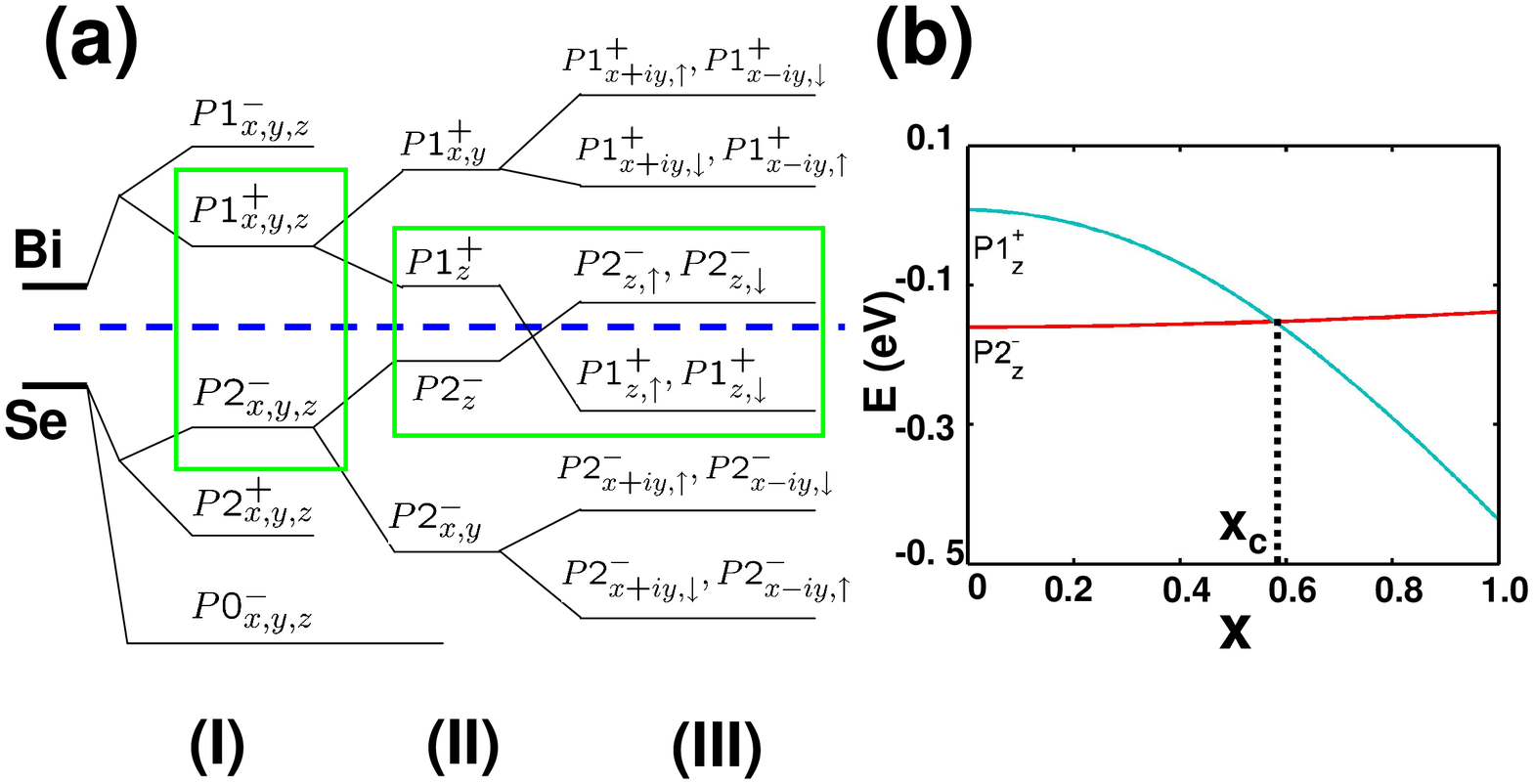}
    \end{center}
    \caption{ {\bf Band Sequence.} (a) Schematic picture of the evolution from the atomic $p_{x,y,z}$ orbitals of $Bi$ and $Se$
    into the conduction and valence bands of $Bi_2Se_3$ at $\Gamma$ point. The three different stages
    (I), (II) and (III) represent the effect of turning on chemical bonding, crystal field splitting
    and spin-orbit coupling, respectively (see text). The blue dashed line
    represents the Fermi energy. (b) The energy levels $|P1^+_z\rangle$ and $|P2^-_z\rangle$ of $Bi_2Se_3$ at $\Gamma$ point versus an artificially rescaled atomic
    spin-orbit coupling $\lambda(Bi)=x\lambda_0(Bi)=1.25x{\rm eV}, \lambda(Se)=x\lambda_0(Se)=0.22x{\rm eV}$ (see text).
    A level crossing occurs between these two states at $x=x_c\simeq 0.6$.
    }
    \label{fig:level}
\end{figure}

{\it Topological surface states}. The existence of topological
surface states is one of the most important properties of the
topological insulators. To see the topological features of the four
systems explicitly, we calculate the surface states of these four
systems based on {\it ab initio} calculation. First we construct the
maximally localized Wanneir function(MLWF) from the {\it ab initio}
calculation using the method developed by N.~Marzari {\it et
al.}\cite{marzari1997,souza2001}. With these MLWF hoping parameters,
we employ iterative method\cite{sancho1984,sancho1985} to obtain the
surface Green function of the semi-infinite system. The imaginary
part of the surface Green function is the local density of states
(LDOS), from which we can obtain the dispersion of the surface
states. When calculating the surface Green function, we only use the
bulk's MLWF hoping parameters as semi-infinite systems without
considering surface corrections. Due to the layered structure of
these materials, we expect the surface reconstruction effect to be
minor for $[111]$ surface. The surface LDOS on the $[111]$ surface
for all the four systems are shown in Fig. 4. For $Sb_2Te_3$,
$Bi_2Se_3$ and $Bi_2Te_3$, one can clearly see the topological
surface states which form a single Dirac cone at $\Gamma$ point. On
comparison $Sb_2Se_3$ has no surface state and is a topological
trivial insulator. Thus the surface state calculation agrees well
with the bulk parity analysis, and confirm conclusively the
topologically nontrivial nature of the three materials. For
$Bi_2Se_3$ the Fermi velocity of the topological surface states is
$v_F\simeq 5.0\times 10^5{\rm m/s}$, which is similar to that of the
other two materials.

\begin{figure}
   \begin{center}
\includegraphics[angle=0,width=3.5in]{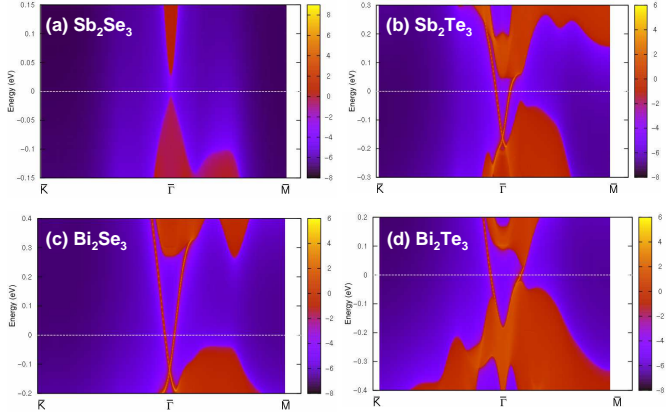}
    \end{center}
    \caption{{\bf Surface states}. Energy and momentum dependence of the local density of states (LDOS) for (a) $Sb_2Se_3$, (b) $Sb_2Te_3$, (c)
    $Bi_2Se_3$ and (d) $Bi_2Te_3$ on the $[111]$ surface. Here warmer color represents higher LDOS. The red regions indicate
    bulk energy bands and the blue regions indicate bulk energy gap.
    The surface states can be clearly seen around $\Gamma$ point as red lines dispersing in the bulk gap
    for $Sb_2Te_3$, $Bi_2Se_3$ and $Bi_2Te_3$. No surface state
    exists
    for $Sb_2Se_3$.
    }
    \label{fig:ss}
\end{figure}

{\it Low energy effective model.} Since the topological nature
is determined by the physics near $\Gamma$ point, it is
possible to write down an simple effective Hamiltonian to
characterize the low-energy long-wavelength properties of the
system.
Starting from four low lying states
$\left|P1^+_z,\uparrow(\downarrow)\right\rangle$ and
$\left|P2^-_z,\uparrow(\downarrow)\right\rangle$ at $\Gamma$ point,
such a Hamiltonian can be constructed by the theory of invariants\cite{winkler2003}
for the finite wavevector ${\bf k}$.
Based on the symmetries of the system,
the generic form of the $4\times4$ effective Hamiltonian
can be written down up to the order of $O({\bf k}^2)$, and
the tunable parameters in the Hamiltonian can be
obtained by fitting the band structure of our {\it ab
initio} calculation.
The important symmetries of the system are
time-reversal symmetry $T$, inversion symmetry $I$ and three fold
rotation symmetry $C_3$ along the $z$ axis. In the basis of
$\left(\left|P1^+_z,\uparrow\right\rangle,
\left|P2^-_z,\uparrow\right\rangle,
\left|P1^+_z,\downarrow\right\rangle,
\left|P2^-_z,\downarrow\right\rangle\right)$,
the representation of the symmetry operations are given by
$T=\mathcal{K}\cdot i\sigma^y\otimes{\rm I}_{2\times 2}$, $I={\rm
I}_{2\times 2}\otimes\tau_3$ and
$C_3=\exp\left(i\frac{\pi}3\sigma^z\otimes{\rm
I}_{2\times2}\right)$, where $\mathcal{K}$ is the complex
conjugation operator, $\sigma^{x,y,z}$ and $\tau^{x,y,z}$ denote the
Pauli matrices in the spin and orbital space, respectively. By
requiring these three symmetries and keeping only the terms up to
quadratic order in ${\bf k}$, we obtain the following generic form
of the effective Hamiltonian:
\begin{eqnarray}
    &&H({\bf k})=\epsilon_0({\bf k}){\rm I}_{4\times 4}+\nonumber\\
    &&\left(
    \begin{array}{cccc}
        \mathcal{M}({\bf k})&A_1k_z&0&A_2k_-\\
        A_1k_z&-\mathcal{M}({\bf k})&A_2k_-&0\\
        0&A_2k_+&\mathcal{M}({\bf k})&-A_1k_z\\
        A_2k_+&0&-A_1k_z&-\mathcal{M}({\bf k})
    \end{array}
    \right)+o({\bf k}^2)
    \label{eq:Heff}
\end{eqnarray}
with $k_\pm=k_x\pm ik_y$, $\epsilon_0({\bf
k})=C+D_1k_z^2+D_2k_\perp^2$ and $\mathcal{M}({\bf
k})=M-B_1k_z^2-B_2k_\perp^2$. By fitting the energy spectrum of the
effective Hamiltonian with that of the {\it ab initio} calculation,
the parameters in the effective model can be determined. For
$Bi_2Se_3$, our fitting leads to $M=0.28eV$, $A_1=2.2eV\cdot$\AA,
$A_2=4.1eV\cdot$\AA, $B_1=10eV\cdot$ \AA$^2$,
$B_2=56.6eV\cdot$\AA$^2$, $C=-0.0068eV$, $D_1=1.3eV\cdot$ \AA$^2$,
$D_2=19.6eV\cdot$\AA$^2$. Except for the identity term
$\epsilon_0({\bf k})$, the Hamiltonian (\ref{eq:Heff}) is nothing
but the 3D Dirac model with uniaxial anisotropy along
$z$ direction and ${\bf k}$ dependent mass terms. From the fact
$M,B_1,B_2>0$ we can see that the order of the bands
$\left|T1^+_z,\uparrow(\downarrow)\right\rangle$ and
$\left|T2^-_z,\uparrow(\downarrow)\right\rangle$ are inverted around
${\bf k}=0$ compared with large ${\bf k}$, which correctly
characterizes the topologically non-trivial nature of the system.
Such an effective Dirac model can be used for further theoretical
study of the $Bi_2Se_3$ system, as long as the low energy properties
are concerned. For example, as one of the most important low energy
properties of the topological insulators, the topological surface
states can be obtained from diagonalizing the effective Hamiltonian
(\ref{eq:Heff}) with an open boundary condition, with the same
method used in the study of two-dimensional quantum spin Hall
insualtor\cite{koenig2008}. For a surface perpendicular to the $z$
direction ({\it i.e.}, $[111]$ direction), the surface states are
described by a $2\times 2$ massless Dirac Hamiltonian
\begin{eqnarray}
H_{\rm
surf}(k_x,k_y)=\left(\begin{array}{cc}0&A_2k_-\\A_2k_+&0\end{array}\right)\label{Hsurf}
\end{eqnarray}
in the basis of $(\left|{\bf k},\uparrow\right\rangle,\left|{\bf
k},\downarrow\right\rangle)$. Here the surface state wavefunction
$\left|{\bf k},\uparrow (\downarrow)\right\rangle$is a superposition
of the $\left|P1_z^+,\uparrow (\downarrow)\right\rangle$ and
$\left|P2_z^+,\uparrow (\downarrow)\right\rangle$, respectively. For
$A_2=4.1{\rm eV}\cdot$\AA  obtained from the fitting, the fermi
velocity of surface states is given by $v_F=A_2/\hbar\simeq
6.2\times 10^5{\rm m/s}$, which agrees reasonably with the {\it
ab initio} results shown in Fig. 4 (c). In summary, the surface
effective theory (\ref{Hsurf}) characterizes the key features of the
topological surface states, and can be used in future to study the
surface state properties of the $Bi_2Se_3$ family of topological
insulators.

In conclusion we have theoretically predicted a new class of
topological insulators $Sb_2Te_3$, $Bi_2Te_3$ and $Bi_2Se_3$. In
particular, $Bi_2Se_3$ has a large topologically non-trivial energy
gap $\sim 0.3{\rm eV}$, sufficient for room temperature operation.
The topologically nontrivial nature of these three materials
originates from a band inversion at $\Gamma$ point, similar to the
strained 3D $HgTe$\cite{dai2008,fu2007a} and two-dimensional $HgTe$
quantum wells\cite{bernevig2006d}. The topologically robust surface
states are studied by {\it ab initio} method, which consist of a
single Dirac cone around the $\Gamma$ point. We have also obtained a
$4\times 4$ effective theory to characterize the bulk properties at
low-energy and long-wavelength, and a $2\times 2$ massless Dirac
model to describe the surface states.

The topological surface states can be directly verified by various
experimental techniques, such as ARPES and scanning tunneling
microscopy (STM). In the recent years, evidences of surface states
have been observed for $Bi_2Se_3$ and $Bi_2Te_3$ in
ARPES\cite{noh2008} and STM\cite{urazhdin2004} experiments. In
particular, the surface states of $Bi_2Te_3$ observed in
Ref.\cite{noh2008} had a similar dispersion as we obtained in Fig. 4
(d), which were also shown to be quite stable and robust, regardless
of photon exposure and temperature. Thus this experimental result
strongly supports that the surface states have topological origin.
Further experimental studies on the surface state properties, such
as ARPES, STM and transport measurements are necessary to verify our
prediction. Moreover, the 3D topological insulators are predicted to
display the universal topological magneto-electric
effect\cite{qi2008} when the surface is coated with a thin magnetic
film. Compared with the $Bi_{1-x}Sb_x$ alloy, the surface states of
the $Bi_2Se_3$ family of topological insulators contain only a
single fermi pocket, making it easier to open up a gap on the
surface by magnetization and to observe the topological Faraday/Kerr
rotation\cite{qi2008} and image magnetic monopole
effect\cite{qi2008monopole}. If observed, such effects can be
unambiguously identified as the experimental signature of the
non-trivial topology of the electronic properties.


\end{document}